\documentclass[11pt,twoside]{article}

\usepackage{asp2004}
\usepackage{epsf}
\usepackage{psfig}
\usepackage{graphicx}
\usepackage{lscape}

\newcommand{\bra}{Br$\alpha$}
\newcommand{\brg}{Br$\gamma$}

\begin{document}
\title{K-band spectroscopy of deeply embedded, young OB stars}    
\author{A. Bik$^{1}$, L. Kaper${^2}$,  M.M. Hanson$^{3}$, L.B.F.M. Waters$^{2,4}$}   
\affil{$^1$European Southern Observatory, Karl-Schwarzschild
           Strasse 2, Garching-bei-M\"unchen, D85748, Germany\\
$^2$Astronomical Institute ``Anton Pannekoek'',
           University of Amsterdam, Kruislaan 403, 1098 SJ Amsterdam,
           The Netherlands\\
$^3$University of Cincinnati, Cincinnati, OH 45221-0011, U.S.A.  \\
$^4$Instituut voor Sterrenkunde, Katholieke Universiteit Leuven,
           Celestijnenlaan 200B, B-3001 Heverlee, Belgium}    

\begin{abstract} 
We have obtained high resolution (R = 10,000) K-band spectra of
   candidate young massive stars deeply embedded in high-mass
   star-forming regions. These objects were selected from a
   near-infrared survey of 44 regions of high-mass star-formation
   (\cite{Kaper}). In these clusters, 38 OB stars are identified whose
   K-band spectra are dominated by photospheric emission.  In almost
   all those stars, the K-band spectra are indistinguishable from
   field stars.  However, in some stars the profile of the Br$\gamma$
   line is different (less deep, or absent) from those of the O field
   stars. One of the explanations of these profiles might be an
   enhanced mass-loss.

\end{abstract}


\section{Introduction}

The advent of high-quality near-infrared instrumentation has opened up
a new window on the birth sites of massive stars. The formation time
scales of a massive star is short, of the order of 100,000 years. This
means that the newly born stars are still deeply embedded in their
parental molecular cloud. Not much is known about the onset of the
stellar wind or the initial rotation properties of young massive
stars. Occurring behind large amounts of extinction, inside their
natal molecular cloud, these evolutionary stages can only be observed
at near-infrared wavelengths.

We have identified 38 OB type stars inside massive star forming
regions (\cite{Ostarpaper}) using K-band spectroscopy. They are
classified based on the classification scheme of
\cite{hanson96}. Three O3-O4 stars are detected, while only a few O3
stars are known in our galaxy. The K-band spectra of the OB stars are
in general similar to those of the (more evolved) OB stars located in
OB associations and the field. This suggest that the K-band spectral
properties of the very young OB stars are already indistinguishable
from those of the more evolved stars.

\section{Stellar winds}

\begin{figure}[!ht]

\includegraphics[width=11cm]{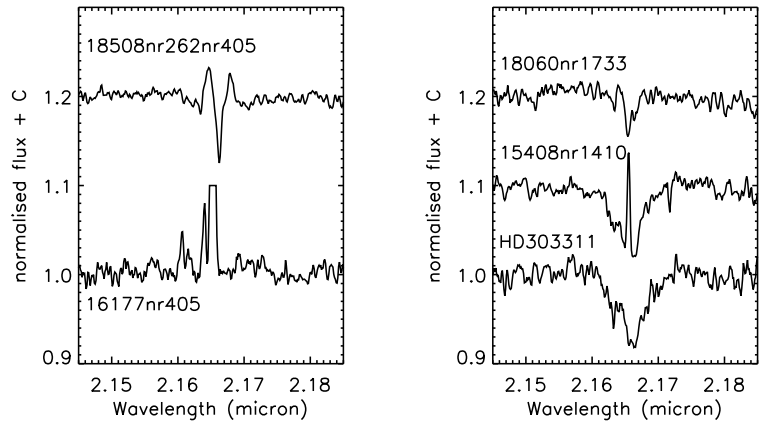}
\caption{\brg\ profile of 4 young O5-O6 V stars in comparison
  with a \brg\ profile of a O5V field star (lower spectrum, right
  panel).}\label{fig:brg}

\end{figure}

A few stars classified as early O stars (O5-O6V) show a peculiar
behaviour of the \brg\ line (Fig. \ref{fig:brg}); the line is either not present, in
emission, or the profile is partly filled in. This might be caused by
an enhanced mass-loss. However, this would only happen if the stellar
mass-loss is as high as that of supergiants (\cite{Lenorzer04},
Lenorzer et al, these proceedings). Veiling by dust or free-free
emission is unlikely as the intrinsically weak lines like {\sc civ}
and {\sc niii} are detected.

Contrary to the suggestion found in our data that the young stars
possess a relatively strong wind, evidence has been presented for a
class of young OB stars which have unusually weak winds for their
spectral type (\cite{Heydari02}, \cite{Martins04}).

To distinguish between these two scenarios for the stellar wind of
the youngest OB stars, a careful analysis of spectral lines sensitive
to the wind density is needed. The \brg\ line in the K-band is
not the best line for this analysis. \bra\ in the L-band is
more sensitive to the stellar wind density (\cite{Lenorzer04}).
Currently a observing program to obtain high resolution spectra of the
\bra\ line with ISAAC on the VLT is underway.





\end{document}